\documentclass[10pt]{iopart}

\usepackage{graphicx}
\usepackage{color}

\newcommand{\tg}{\ensuremath{t_{2g}}}
\newcommand{\eg}{\ensuremath{e_{g}}}
\newcommand{\EF}{\ensuremath{E_{F}}}
\newcommand{\muB}{\ensuremath{\mu_{B}}}
\newcommand{\aeq}{\ensuremath{a_{\mathrm{eq}}}}

\begin{document}

\title[Properties of tetrahedrally bonded half-metals]{A review of the
electronic and magnetic properties of tetrahedrally bonded
half-metallic ferromagnets}

\author{Ph Mavropoulos\dag\ and I Galanakis\ddag}

\address{\dag\ Institut f\"ur Festk\"orperforschung, Forschungszentrum J\"ulich, D-52425
J\"ulich, Germany}
\address{\ddag\ Department of Materials Science, School of Natural
  Sciences, University of Patras, Patras 265 04, Greece}

 \ead{ph.mavropoulos@fz-juelich.de, galanakis@upatras.gr}

\begin{abstract}
The emergence of the field of spintronics brought half-metallic
ferromagnets to the center of scientific research. A lot of interest
was focused on newly created transition-metal pnictides (such as CrAs)
and chalcogenides (such as CrTe) in the metastable zinc-blende lattice
structure. These compounds were found to present the advantage of high
Curie temperature values in addition to their structural similarity to
semiconductors. Significant theoretical activity has been devoted to
the study of the electronic and magnetic properties of these compounds
in an effort to achieve a better control of their experimental
behaviour in realistic applications. This review is devoted to an
overview of the studies of these compounds, with emphasis on
theoretical results, covering their bulk properties (electronic
structure, magnetism, stability of the zinc-blende phase, stability of
ferromagnetism) as well as low-dimensional structures (surfaces,
interfaces, nanodots and transition-metal delta-doped semiconductors)
and phenomena that can possibly destroy the half-metallic property,
like structural distortions or defects.
\end{abstract}



\maketitle

\tableofcontents

\section{Introduction\label{sec:1}}

The rapid advance of magnetoelectronics and spintronics in recent
years has given a strong boost in the search for novel magnetic
materials, as are the half-metallic ferromagnets. These are spin
polarized materials, which exhibit the property of having a metallic
density of electron states for the one spin direction (usually
majority spin), while there is a band gap around the Fermi level,
\EF, for the states of the opposite spin. This \emph{half-metallic}
property (the name coined by de Groot and collaborators
\cite{deGroot83}) makes the one spin channel conducting, while the
other is insulating (or semiconducting), so that, in the ideal case,
half-metals can conduct a current which is 100\% spin polarized.

The spin polarization $P$ at $E_F$ is expressed in terms of the
spin-up and spin-down density of states, $n_{\uparrow}(E_F)$ and
$n_{\downarrow}(E_F)$, as
\begin{equation}
P =
\frac{n_{\uparrow}(E_F)-n_{\downarrow}(E_F)}{n_{\uparrow}(E_F)+n_{\downarrow}(E_F)}.
\label{eq:1}
\end{equation}
For an ideal half-metallic ferromagnet, $P=1$; however, a number of
factors can reduce $P$ to lower values: defects and disorder,
spin-orbit coupling, temperature, interface and surface states in the
gap, etc.  Apart from a polarization value as close as possible to
unity, half-metals must have also other properties in order to be
functional: a high Curie point $T_C$ and compatibility with
semiconductors are among the key points for spintronics applications.

In 2000, Akinaga and collaborators \cite{Akinaga00} reported growth of
a few layers of CrAs in the metastable zinc-blende phase (zb) for the
first time; this was achieved by molecular beam epitaxy on a GaAs
substrate (the ground state structure of CrAs is the MnP
structure~\cite{Watanabe69}).  In the zb phase, CrAs was reported to be
ferromagnetic, with a Curie point higher than 400~K. In the same
publication, first-principles calculations of bulk zinc-blende CrAs
showed the material to be a half-metallic ferromagnet; the moment per
formula unit was calculated to be 3~$\mu_B$, in agreement with the
experimental result.

These findings initiated a strong activity, because several merits
were combined: the half-metallic property, coherent growth on a
semiconductor, and $T_C$ higher than room temperature. The activity
was extended beyond zb CrAs, encompassing a variety of tetrahedrally
bonded transition metal (TM) compounds with $sp$ atoms of the IV, V
and VI groups of the periodic table. In this paper we review relevant
investigations up to date, with emphasis on the theoretical results on
the electronic structure and magnetic properties of these compounds.
We do not extend our review to the related but vast topics of diluted
magnetic semiconductors or other types of half-metals.  We also note
in passing the surprising (calculated) appearance of half-metallic
ferromagnetism in tetrahedrally bonded compounds of I-V and II-V
elements~\cite{Kusakabe04,Sieberer06,Yao06}, such as zb CaAs, where no
transition metals are involved. The origin of ferromagnetism is
therefore of a different nature here, but a discussion of these
compounds is beyond the purpose of the present work.

The remainder of the paper is summarized as follows. We begin with
a summary of experimental findings in section \ref{sec:2}. Section
\ref{sec:3} is devoted to an analysis of the electronic and
magnetic structure of half-metallic zinc-blende compounds, as
obtained by first-principles calculations.  Investigations on the
magnetic ground state (ferromagnetic vs. antiferromagnetic),
exchange interactions and Curie temperatures of such compounds are
reviewed in section \ref{sec:Ferro}. The metastability of the
zinc-blende phase, compared to the NiAs or other phases, is
addressed in section \ref{sec:6}. The possibility of surface
half-metallicity is discussed in section \ref{sec:4}.  In section
\ref{sec:5} we review the properties of the interfaces with
semiconductors, including delta-doping and transport in
multilayered structures.  We continue with a discussion on the
effect of disorder on the half-metallic gap in section
\ref{sec:7}. Section \ref{sec:8} is devoted to tetrahedrally
bonded TM-$sp$ compounds with other structures, such as the
wurtzite structure. Finally, in section \ref{sec:12} review
results of calculations beyond density-functional theory. We
conclude with a summary in section~\ref{sec:13}.

\section{Experimental results\label{sec:2}}

Experimentally, CrAs in the zinc-blende structure was first grown
by Akinaga {\it et al.}~\cite{Akinaga00}, by molecular beam
epitaxy (MBE) on a GaAs substrate. The zb phase could be
stabilized only for thin CrAs layers; for samples above a critical
thickness of 3~nm, the reflection high-energy electron diffraction
(RHEED) pattern indicated an unknown different phase; such small
thicknesses are typical in all experiments involving zb pnictides
(compounds including group-V atoms) or chalcogenides (compounds
including group-VI atoms). The measured saturation magnetization
at low temperature corresponded to 3~\muB\ per formula unit, in
good agreement with {\it ab-initio} calculations (which also
predicted a half-metallic band structure). The precise Curie
temperature was not reported, but it was found to be higher than
400~K, suggesting applicability at room temperature. The surface
morphology and magnetic characteristics of zb CrAs films under
different growth conditions were studied in
reference~\cite{Mizuguchi02}. In reference~\cite{Mizuguchi02c}, a
thickness-dependence angular resolved photoemission spectroscopy
(ARPES) study showed a dispersion for a 2~nm CrAs film, which was
not present for a 30~nm film; this, together with the RHEED
pattern, suggested a monocrystalline sample at 2~nm, changing to
polycrystalline at 30~nm. ARPES data on a thin zb-CrAs film were
also presented in reference~\cite{Oshima02}.

Further work~\cite{Ofuchi03}, based on fluorescence extended X-ray
absorption fine structure (EXAFS), provided more evidence on the
existence of zb phase on a 2~nm zb CrAs film, and found a CrAs
bond length of 2.49~\AA\ (in close agreement with the value of
2.52~\AA\ which is deduced from {\it ab-initio}
calculations~\cite{Shirai01}). Moreover, high-resolution
transmission electron micrography (HRTEM) and X-ray grazing
incidence diffraction (GID) were used in reference~\cite{Bi06} to
identify 2.5~nm ($~9$ monolayers (ML)) thick zb CrAs on GaAs; it
was found that the zb-phase can be partly deformed by appropriate
annealing. The samples were ferromagnetic, the saturation magnetic
moment of both phases was found to be about 3~\muB\ per formula
unit, and $T_C$ was above 400~K.

In parallel, zb CrAs/GaAs multilayers were grown by use of MBE and
studied in references~\cite{Mizuguchi02b} and \cite{Akinaga04}. In
both studies the optimal (for the zb-structure) multilayers
consisted of a period of 2~ML of GaAs and 2~ML of CrAs. The period
was repeated 20 times in \cite{Mizuguchi02b} and 100 times in
\cite{Akinaga04}. Magnetization measurements in \cite{Akinaga04}
showed a magnetic moment of 2~\muB\ per formula unit of CrAs
(lower than the theoretical prediction of 3~\muB), and a Curie
temperature of 800~K.

CrSb was also grown by MBE in ultrathin films of a few monolayers
in the zinc-blende structure~\cite{Zhao01,Deng06b}. In the work of
reference~\cite{Zhao01}, a GaAs, GaSb, and (Al,Ga)Sb substrates
were used. The structure was examined by HRTEM and RHEED. The
samples were ferromagnetic with a magnetic moment per formula unit
reported to be between 3 and 5~\muB\ (the theoretical prediction
is 3~\muB\ \cite{Shirai98,Liu03,Galanakis03}) and a $T_C$
definitely exceeding 400~K. In reference~\cite{Deng06b} it was
shown how the substrate can be tailored for growth of somewhat
thicker films of CrSb. While the maximal thickness on GaAs was
found to be 1~nm (3~ML), an (In,Ga)As substrate provides a larger
lattice parameter which can be tuned by changing the In
composition to favor the growth of zb-CrSb. On an
In$_{0.08}$Ga$_{0.92}$As substrate, 6~ML of zb-CrSb were
successfully grown, and for an In$_{0.13}$Ga$_{0.87}$As substrate,
growth of a 9~ML (3~nm) zb-CrSb film could be achieved.

Also CrSb/GaAs multilayers in the zb structure were
grown~\cite{Zhao03}. Here, the CrSb thickness was just 1~ML, while the
GaAs layers were 5~nm thick, and the multilayers included up to four
periods. The zb-structure was monitored by RHEED measurements during
growth, and cross-checked for dislocations at the interfaces by
HRTEM. The saturation magnetization at low temperatures was close to
the theoretical value of 3~\muB. The ferromagnetic transition
temperature was reported to be above 400~K (measurements at high
temperatures were limited by the superconducting quantum interference
device (SQUID)).

MnAs could be grown in nanodots in the zinc-blende
structure~\cite{Oshima02,Ono02,Okabayashi04}\footnote{In
\cite{Oshima02} it is reported that MnSb dots could not be
directly grown in the zb-phase on GaAs, probably due to lattice
mismatch.}. References~\cite{Oshima02} and \cite{Ono02} report
growth of zb-MnAs dots of average size of about 16~nm on
sulphur-terminated GaAs(001) surfaces, with a density of dots of
$2.3\times 10^{10}$/cm$^2$. When the growth was continued, the
MnAs dots changed to the NiAs structure (which is the equilibrium
structure of MnAs). SQUID measurements showed the zb-MnAs dots to
be ferromagnetic, with a Curie temperature of about 280~K.
Photoemission spectra~\cite{Oshima02,Ono02} show a localization of
the Mn $d$ states, with absence of a Fermi edge\footnote{This
absence of a Fermi edge, indicating an insulating phase, is not
found by {\it ab-initio} calculations. In \cite{Oshima02,Ono02} it
is speculated that the reason is strong electron correlations,
which are not well captured by local density-functional theory.},
a main peak at around $4$~eV below \EF, and a satellite at $6$~eV
below \EF; these features are similar to (Ga,Mn)As diluted
magnetic semiconductors, and quite different than MnAs in the NiAs
structure, where a Fermi edge is clearly seen in
photoemission~\cite{Oshima02,Ono02}. Similar are the conclusions
of reference~\cite{Oshima02}, where it is reported that, with
increasing dot density (above $3\times 10^{12}$/cm$^2$), a Fermi
edge evolves in the photoemission spectrum. This onset of metallic
behaviour is attributed by the Authors either to formation of a
NiAs-type of structure or to percolation among the clusters. We
also note the possibility of delta-doped GaAs with Mn
\cite{Kawakami00,Tanaka05}; see section \ref{sec:delta} for
further discussion.

Further experimental work on zinc-blende pnictides and chalcogenides
includes MBE growth of GaSb(25 \AA)/MnSb(2 \AA) (001)
multilayers~\cite{Choi05} and CrTe on GaAs(001) thin
films~\cite{Sreenivasan06}. In both cases it is reported that there
are strong indications of a zb phase, but no definite proof.

Finally, we note that, to our knowledge, the half-metallic property
itself has not been experimentally proven or disproven in these
compounds.

\section{Calculations on TM-based zinc-blende compounds\label{sec:3}}

\subsection{Band structure and DOS}

\begin{figure}
\begin{center}
\includegraphics[angle=270,width=0.7\linewidth]{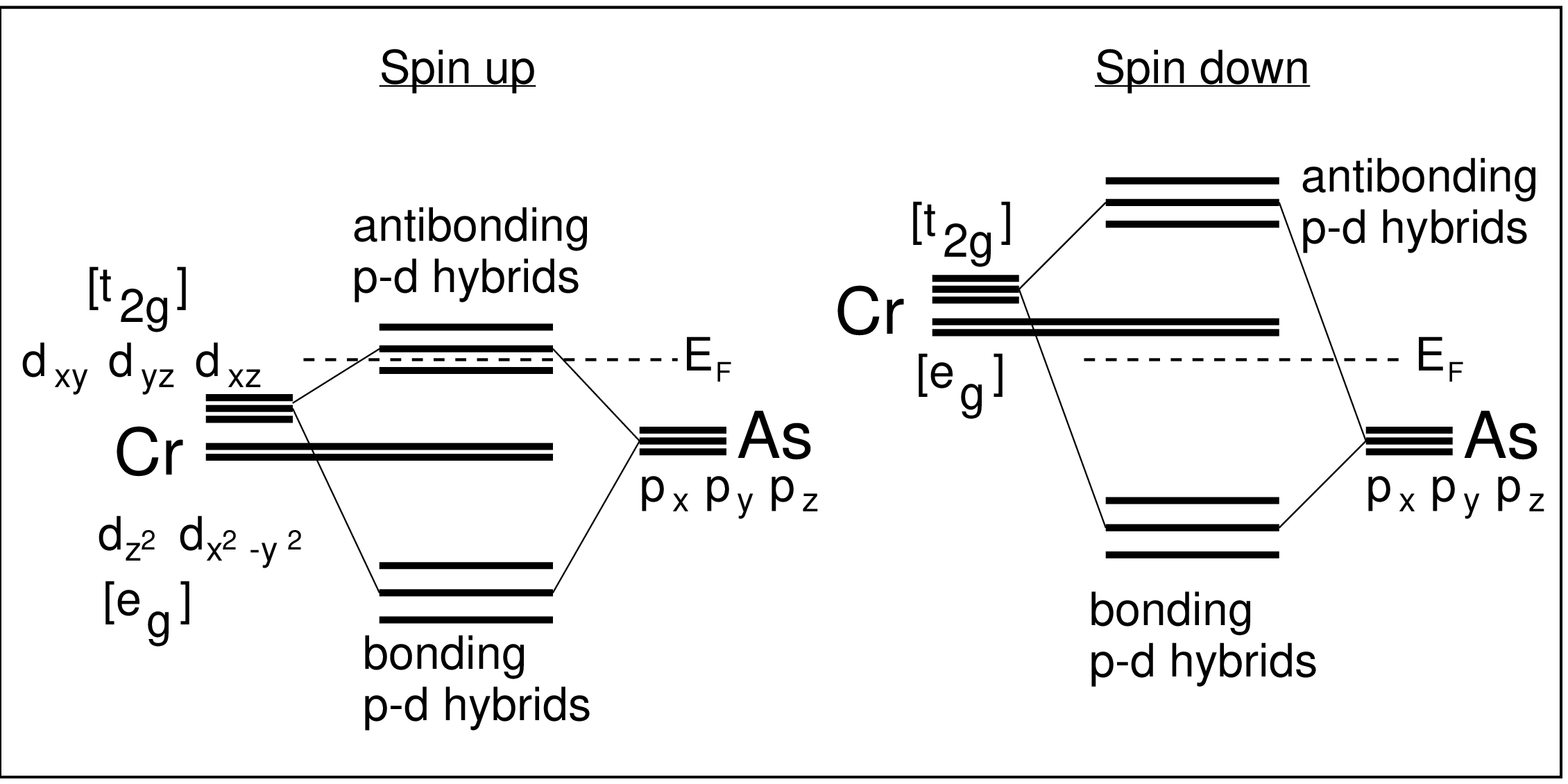}
\includegraphics[width=0.8\linewidth]{bands_dos_CrAs.eps}
\caption{Up: Schematic description of the $p$-$d$ hybridization leading to
  the half-metallic property. Down: Calculated spin-resolved band
  structure and spin- and atom-resolved density of states of
  zinc-blende CrAs in the GaAs lattice parameter within the LSDA.
\label{fig:1}}
\end{center}
\end{figure}

The principal mechanism leading to the appearance of the half-metallic
gap in zinc-blende compounds of transition metals (TM) with $sp$
elements is the hybridization of the $d$ wavefunctions of the TM with
the $p$ wavefunctions of the $sp$ atom. This \emph{$p$-$d$
hybridization} is dictated by the tetrahedral environment, where each
atom is surrounded by four atoms of the other species. In particular,
in the presence of tetrahedral symmetry the $d$-states split in two
irreducible subspaces: the threefold-degenerate \tg\ subspace,
consisting of $d_{xy}$, $d_{yz}$ and $d_{xz}$ states, and the
twofold-degenerate \eg\ subspace, which includes the $d_{x^2-y^2}$ and
$d_{z^2}$ states. Only states of the former subspace can hybridize
with the $p$ states of the $sp$-atom neighbours, forming bonding and
antibonding hybrids; the \eg\ states, on the contrary, remain rather
non-bonding.  The situation is shown schematically in figure
\ref{fig:1} (up) for both spins, with the majority (spin-up) $d$
states being lower in energy than the spin-down states due to the
exchange splitting.

Starting from this picture, we can proceed now to a discussion of the
energy bands, shown in figure \ref{fig:1} (down), and calculated
within the local spin density approximation (LSDA) of
density-functional theory (the As $4s$ states are much lower in energy
and omitted in the figure). We note that the notations \eg\ and \tg\
are, strictly speaking, valid only for states at the center of the
Brillouin zone; however, the energy bands formed by the \eg\ and \tg\
states are energetically rather separated, therefore we keep this
notation for the bands formed by these states.  The strong $p$-$d$
hybridization leads to wide corresponding \tg\ energy bands, located,
for spin up, at about $-5$ to $-2$ eV (bonding states), and above \EF\
(antibonding); the latter are in the same energy range with the Cr $s$
states. The rather non-bonding \eg\ states, on the contrary, form
mainly narrow bands. These remain energetically between the bonding
and antibonding $p$-$d$ bands, at about $-1$ eV. For spin down, the
situation is similar, with all bands shifted to higher energies
because of the exchange splitting; the spin-down bands are also wider,
because they reside at higher energies.  The striking effect is that,
for spin-down, the exchange splitting is strong enough to push the
\eg\ states just above the Fermi level, so that \EF\ is in the gap.
This is also seen in the atom- and spin-resolved density of states
(DOS).  According to the calculated DOS, the material is manifestly
half-metallic in its electronic and magnetic ground state.

This basic form of the electronic density of states is found by all
relevant electronic structure calculations, within either the LSDA or
the generalized gradient approximation (GGA) to density-functional
theory.  Calculations within the GGA show a slightly stronger local
magnetic moment at the Cr site, resulting in a stronger exchange
splitting and a wider gap; then, \EF\ is even deeper within the
gap. It is generally expected that the GGA yields a more accurate
prediction of the lattice parameter and  magnetic moment in
magnetic compounds, with corrections of the order of 2-3\%.

The $p$-$d$ hybridization, essential to the formation of the gap, is
well-known as \emph{p-d repulsion} in the physics of TM defects in
semiconductors of zinc-blende (where the TM substitutes the cation) or
diamond structure~\cite{Wei87}, and is essentially present also in
diluted magnetic semiconductors (DMS) \cite{Akai98}. In this respect,
the electronic structure of zinc-blende pnictides and chalcogenides
can be viewed as the electronic structure of TM-doped semiconductors
in the limit of high concentration. From the electronic structure
point of view, it is expected that all zinc-blende binary compounds of
TM with group IV, V, and VI $sp$-atoms are candidates for the
appearance of half-metallic ferromagnetism---in fact, their calculated
density of states is similar to the one of CrAs~(e.g. see references
\cite{Akinaga00,Galanakis03,Zhang04b,Yao05,Sakuma02,Xu02,Shirai03,Pask03,Xie03,Xie03b,Sanyal03,Zhang03,Zhang04}),
with spin-down gaps around 0.5--1~eV when the $sp$ element belongs to
the IV-group, around 1--2~eV when it belongs to the V-group, and
1.5--2.5~eV when it belongs to the VI-group, and with the spin-up
bands appropriately shifted to achieve charge neutrality.

The half-metallic state is, however, not guaranteed in all of
these compounds, nor is a ferromagnetic ground state. We discuss
this in sections \ref{sec:Ferro} and \ref{sec:6}.  As we shall
see, other factors (lattice constant, antiferromagnetic
interactions, structural instabilities, etc.) limit this
expectation, but a number of compounds remains within these
limits.

\subsection{Magnetic moments}

In the bulk of half-metallic compounds, the spin moment per formula
unit is an integer (in \muB). Since \EF\ is in the gap, the total
number of spin-down valence electrons per formula unit,
$N_{\mathrm{tot}}^{\downarrow}$, is an integer. The total number of
valence electrons per formula unit, $Z_{\mathrm{tot}}$, must also be
an integer, so it follows that the number of valence spin-up
electrons, $N_{\mathrm{tot}}^{\uparrow} = Z_{\mathrm{tot}} -
N_{\mathrm{tot}}^{\downarrow}$, is an integer. Thus the magnetic
moment, $M = (N_{\mathrm{tot}}^{\uparrow} -
N_{\mathrm{tot}}^{\downarrow})\muB = (Z_{\mathrm{tot}} - 2
N_{\mathrm{tot}}^{\downarrow})\,\muB$, is an integer. This is a
necessary, but not sufficient, condition for half-metallicity. (At
surfaces or interfaces this can change; see sections~\ref{sec:4} and
\ref{sec:5}.)

Due to the similarity in the DOS of all zinc-blende pnictides and
chalcogenides, the number of spin down valence states up to
the gap is always four: there is the low-lying $s$ band of the $sp$
atom, and the three bonding $p$-$d$ (\tg) bands. Therefore, assuming
half-metallic behaviour, the total moment per formula unit follows a
Slater-Pauling-like ``rule of 8'':
\begin{equation}
M_{\mathrm{tot}} = (Z_{\mathrm{tot}} - 8)\,\muB.
\label{eq:2}
\end{equation}
Analogous relations hold for the half-metallic Heusler alloys,
where there is a Slater-Pauling rule of 18 (in half-Heusler
alloys) and rule of 24 (in full-Heusler alloys)
\cite{Galanakis02a}.  Thus, in the half-metallic state, CrAs and
CrSb have a moment of 3~\muB, while CrSe or MnAs (having one more
valence electron) show a moment of 4~\muB. Once more an analogy to
the diluted magnetic semiconductors can be seen: calculations show
that, for instance in Ga$_{1-x}$Mn$_x$As, the spin moment per Mn
atom is exactly 4~\muB. Note that also ternary compounds, such as
(Cr,Mn)As, follow the same rules, as long as all TM (here Cr and
Mn) occupy the same sublattice of the zinc-blende structure; for
instance, see reference~\cite{Fong04} for calculations on
half-metallic CrAs/MnAs(001) superlattices.

From an alternative viewpoint, we can say that the $p$ states,
low-lying as they are in energy, act as reservoirs, hosting electrons
which are donated by the TM until the $p$ states are filled up with 6
electrons. The remaining TM $d$ electrons build up the spin
moment. This viewpoint neglects the essential $p$-$d$ hybridization,
but is useful for electron counting.

Similar to the pnictides and chalcogenides is the electronic
structure of zb TM-compounds with Si, Ge, and Sn
\cite{Sakuma02,Sasioglu05,Liu06,Qian06}. However, zb TM-compounds
with C (e.g. MnC) have a distinctly different density of states,
with the half-metallic gap being among the spin-up states, while
the spin-down bands are metallic \cite{Qian04b,Dag05}. This gives
rise to a ``rule of 12'' instead of eq.~(\ref{eq:2}), and
originates from the huge gap induced by carbon \cite{Pask03}.

\subsection{Local moment trends with cation valence and electronegativity}

In terms of the $p$-$d$ hybridization one can understand the DOS
and local magnetic moment trends with changing cation valence
and/or electronegativity. Firstly we note that the $sp$ atom has a
moment oppositely oriented to the one of the TM; calculated values
are shown in table~\ref{table:1} (up). To understand this, we
observe that the spin-up $d$ states of the TM are lower in energy
than the spin-down $d$ states; thus, when the $p$-$d$
hybridization occurs, the bonding spin-up \tg\ hybrids have a
larger $d$ admixture and are more itinerant than the spin-down
ones; the latter are more localized within the $sp$ atom. Thus, in
the Wigner-Seitz cell of the $sp$ atom, the spin-down states are
in the majority, leading to this ``antiferromagnetic'' coupling.

According to this picture, the following trends can be deduced under
unchanged lattice parameter. On changing the $sp$ atom for a lighter
one of the same group (therefore more electronegative), the $p$ states
are shifted to lower energies and the spin-down $p$-$d$ hybridization
is reduced; the spin-down bonding $p$-$d$ bands become more localized
at the $sp$ atom, transferring to its vicinity spin-down charge from
the TM. On the other hand, because of the proximity of the spin-up $d$
states, the spin-up hybridization is reduced to a lesser extent (this
can be seen by a tight-binding argument, as the hopping decreases with
the energy separation). As a result, the local moment of the $sp$ atom
increases; the same is true for the local moment of the TM, since the
sum, $M_{\mathrm{tot}}$, must be constant integer. Also the gap
increases for lighter $sp$ atoms, because the $p$ states (forming the
valence band) are shifted lower, and the \eg\ states (forming the
conduction band) are shifted higher by the stronger local TM moment
and exchange splitting\footnote{The exchange splitting is roughly
$\sim I\cdot M$, where $M$ is the magnetic moment and $I$ is the
exchange integral, having a value of about 0.9~\muB/eV for transition
metals.}. Note also that TM-group~VI compounds show a wider gap than
TM-group~V compounds, because the $p$ states of the former are lower
in energy.

Competing with these effects is the strong reduction in lattice
parameter due to the smaller volume of lighter $sp$ elements. This
results in reduction of the local moments (due to stronger
hybridization), and also loss of half-metallicity (as we discuss
below); in table~\ref{table:1} (down) we show the effect of lattice
compression on the local moments of CrAs. As it turns out, the effects
of the change of lattice parameter by far outweigh the effects of
electronegativity.

\begin{table}
\begin{tabular}{rrrrrrrrrrr}
\hline
Compound &  CrN  & CrP   &  CrAs &  CrSb &  MnAs &  CrS  &
CrSe &  CrTe \\
\hline
$M$(TM)  &  3.89 & 3.32  & 3.27  & 3.15  & 4.07  & 3.86  &
 3.83  &  3.75  \\
$M(sp)$  &$-1.07$&$-0.45$&$-0.38$&$-0.25$&$-0.25$&$-0.12$&
$-0.10$&$-0.06$ \\
\hline
\end{tabular}
\begin{tabular}{rrrrrr}
\hline
a$_{\mathrm{latt}}$ (\AA) & 5.45 & 5.65 & 5.87 & 6.06 & 6.48 \\
\hline
$M$(Cr) (\muB)            & 2.59 & 3.02 & 3.15 & 3.27 & 3.55 \\
$M$(As) (\muB)            &-0.13 &-0.20 &-0.29 &-0.38 &-0.60 \\
\hline
\end{tabular}
\caption{Up: Local moments (in \muB) of the TM and $sp$ atoms in
  zinc-blende pnictides and chalcogenides calculated within the LSDA
  in the (large) InAs lattice parameter (after
  reference~\cite{Galanakis03}), showing trends due to electronegativity of
  the anion. Down: Effect of changing the lattice parameter on the
  local magnetic moments of CrAs. The magnetization of the
  interstitial region (amounting to about half of the volume in the zb
  structure) is small and not included in the table.\label{table:1}}
\end{table}

\subsection{Effect of lattice parameter on half-metallicity}

We now turn to the effect of the lattice parameter on the
half-metallic property. We begin by the observation that compression
of the lattice shifts all energy levels higher, but has a greater
effect on the extended $p$ states (and $p$-$d$ hybrids), and a lesser
effect on the localized \eg\ states. Since \EF\ is situated among the
spin-up antibonding \tg\ bands, under lattice compression \EF\ will be
carried to higher energies along with these levels; the position of
the spin-down conduction band (which is of \eg\ character) is less
affected by compression. Therefore \EF\ ultimately enters the \eg\
conduction band and half-metallicity is lost. It also turns out that,
under lattice expansion, \EF\ enters the spin-down valence band, also
causing loss of half-metallicity.

As a conclusion, zinc-blende chalcogenides and pnictides are
half-metallic in a limited range of lattice parameters, and it is not
guaranteed that the equilibrium lattice parameter \aeq\ is in this
range. First principles calculations are needed in order to predict
\aeq\ and the magnetic state and DOS at \aeq. For practical purposes,
a deviation from the calculated \aeq\ by a few percent is acceptable,
allowing for a small error in the estimation of \aeq\ by
density-functional calculations\footnote{Especially the LSDA is known
to underestimate, in general, \aeq\ by 2-3\%; GGA corrects this to a
great extent, but in some systems it slightly can overestimate \aeq.},
and for a slightly different lattice parameter of the substrate, on
which the compound can grow pseudomorphically. We remind the reader
that chalcogenides and pnictides are not at equilibrium in the
zinc-blende structure, thus they can be grown only for a few
monolayers on suitable substrates.

In table~\ref{table:2} we summarize results of LSDA
calculations~\cite{Galanakis03} of the half-metallic property at
the lattice parameters of several semiconductors, together with
the equilibrium lattice parameter, for several systems. Systems
marked with a ``$+$'' are half-metallic, with a ``$-$'' are not
half-metallic, while a ``$\pm$'' means that the system is at the
edge of half-metallicity, with \EF\ just touching the conduction
band. Allowing for an LSDA underestimation of the lattice
parameter, together with a small lattice mismatch with the
semiconductor, we identify which systems can, in principle, be
coherently grown (only for a few monolayers; see
section~\ref{sec:6}) on which semiconductors, being at the same
time half-metallic; we mark them with a box. In this way, we
identify the following candidate combinations: VAs on GaAs, VSb on
InAs and GaSb, CrAs on GaAs, CrSb on InAs and GaSb, MnSb on GaSb,
VTe on ZnTe, CrSe on ZnSe and CdS, and CrTe on CdSe and ZnTe. Note
that nitrides, phosphides and sulphides have very small lattice
parameters and are not half-metallic; they are excluded from the
table. Mn-group-VI compounds are not ferromagnetic, and also
excluded.

\begin{table}
\begin{tabular}{rcccccccc}
 a(\AA )& GaN & InN & GaP  & GaAs & InP & InAs & GaSb & InSb \\
 Compound & 4.51 & 4.98 & 5.45 & 5.65 & 5.87 & 6.06 & 6.10 & 6.48 \\
\hline
\hline
(5.54)\hfill VAs & $-$ & $-$ & $-$ & \framebox{$+$} & $+$ & $+$ & $+$ & $+$ \\
(5.98)\hfill VSb & $-$ & $-$ & $-$ & $-$ & $+$ & \framebox{$+$} & \framebox{$+$} & $+$ \\
\hline
(5.52)\hfill CrAs& $-$ & $-$ & $-$ & \framebox{$+$} & $+$ & $+$ & $+$ & $+$ \\
(5.92)\hfill CrSb& $-$ & $-$ & $-$ & $-$ & $+$ & \framebox{$+$} & \framebox{$+$} & $+$ \\
\hline
(5.36)\hfill MnAs& $-$ & $-$ & $-$ & $-$ & $+$ & $+$ & $+$ & $+$ \\
(5.88)\hfill MnSb& $-$ & $-$ & $-$ & $-$ & $-$ & $-$ & \framebox{$\pm$} & $+$ \\
\hline
\hline
\end{tabular}
\begin{tabular}{rcccccc}
 a(\AA )& ZnS & ZnSe & CdS  & CdSe & ZnTe & CdTe \\
 Compound & 5.41 & 5.67 & 5.82 & 6.05 & 6.10 & 6.49 \\
\hline
\hline
(5.56)\hfill VSe  & $-$ & $-$  & $-$  & $+$ & $+$  & $+$ \\
(6.06)\hfill VTe  & $-$ & $-$  & $-$  & $-$ & \framebox{$\pm$}& $+$ \\
\hline
(5.61)\hfill CrSe & $-$ & \framebox{$\pm$}& \framebox{$+$}  & $+$ & $+$  & $+$ \\
(6.07)\hfill CrTe & $-$ & $-$  & $-$  & \framebox{$+$} & \framebox{$+$}  & $+$ \\
\hline
\hline
\end{tabular}
\caption{Half-metallic property of several zinc-blende pnictides and
  chalcogenides at the experimental lattice parameters of several
  semiconductors. The calculated (LSDA) equilibrium lattice parameters
  are given in parentheses. Systems marked with a ``$+$'' are
  half-metallic, with a ``$-$'' are not half-metallic, while a
  ``$\pm$'' means that the system is at the edge of
  half-metallicity. A box indicates that a half-metallic system has a
  suitable lattice parameter to grow on the corresponding
  semiconductor. \label{table:2}}
\end{table}


\subsection{Effect of spin-orbit coupling on the polarization}

Spin-orbit coupling introduces a mixing of the two spin channels, so
that the electron spin is not a good quantum number. Therefore, it is
presence, it is expected that states which were previously accounted
as spin-up states in the region of the half-metallic gap, will now
have some contribution to the spin-down DOS; strictly speaking, the
half-metallic property ($P=100\%$) can never be achieved, since some
amount of spin-orbit coupling is present in all materials. Then, the
half-metallic gap turns into a pseudo-gap, with some spectral
intensity.

However, first-principles calculations \cite{Mavropoulos04,Shirai03b}
show that the effect can be small, of the order of 1\%, except in the
case when heavy elements are present. Thus, when spin-orbit coupling
is included in the calculations, CrAs shows a polarization of
$P=99.6\%$, CrSb shows $P=98.6\%$ (note that Sb is heavier than As),
and MnBi shows a much reduced polarization of $P=77\%$ (neglecting
spin-orbit coupling, MnBi is predicted~\cite{Xu02,Zheng04} to be
half-metallic). These results demonstrate that spin-orbit coupling is
significant for a heavy element such as Bi, also because of its $6p$
states which are directly involved in the band structure around
\EF. The calculated orbital moment of MnBi is large, too, reaching a
value of 0.11~\muB.

A theoretical analysis based on perturbation theory
\cite{Mavropoulos04} shows that the spectral properties in the
half-metallic spin-down pseudo-gap depend (i) on the spin-up DOS in
the same region and (ii) on the proximity of the energy to the band
edges.  It is deduced (and verified by {\it ab-initio} calculations)
that (i) the spin-down DOS is a weak reflection of the spin-up DOS,
depending quadratically on the spin-orbit coupling spin-flip strength,
$n_{\downarrow}(E)\sim |V_{\mathrm{so}}|^2 n_{\uparrow}(E)$; and (ii)
that, close to the band edges, $n_{\downarrow}(E)$ increases strongly,
so that it must be treated beyond first order in perturbation theory
(because of the degeneracy of unperturbed spin-up and spin-down
eigenvalues at the band edge).

\section{Magnetic ground state, exchange interactions, and Curie temperature\label{sec:Ferro}}

In the discussion up to now we assumed a ferromagnetic ground
state, and experiment has shown that CrAs, CrSb, and MnAs in the
zinc-blende phase are ferromagnetic (see section \ref{sec:2}).
However, the ferromagnetic ground state has to be attested by
calculations for the not-yet-fabricated compounds. In a number of
papers
\cite{Shirai01,Wei87,Sakuma02,Shirai03,Sanyal03,Sasioglu05,Zhao02,Kubler03,Kim04,Miao05b,Nakamura05,Nakamura06,Sanyal06}
this aspect is examined via first-principles calculations. In the
framework of such calculations there are three ways to attest the
ferromagnetic ground state. (i) Non-collinear magnetic
calculations can be made, where the system is allowed to relax in
any arbitrary non-collinear configuration. (ii) Total energy
calculations can be made in the ferromagnetic and in several
antiferromagnetic states (perhaps including disordered local
moment states within the coherent potential approximation),
seeking the energy minimum.  (iii) Starting from a ferromagnetic
state, the interatomic exchange constants can be calculated
(usually assuming interactions within a Heisenberg model), from
which conclusions can be drawn as regards the ground state, as
well as excited state properties, including the Curie temperature
(see section \ref{sec:exc}). These methods are to a great extent
complementary, and their applicability depends on the size of the
system as well as the computational method used. The most accurate
way of calculation is probably (i), but it is often not practical
because it requires use of large supercells, solution of equations
coupling the two spin channels, and must be applied to many
starting configurations, especially if there are more than one
local energy minima. Thus, (ii) and (iii) are most common in
practice, and can in most cases ascertain if the ground state is
indeed ferromagnetic or not. We note, however, that if the
magnetic ground state is not ferromagnetic, it is likely to be
non-collinear, since the fcc geometry (corresponding to the TM
sublattice) causes magnetic frustration in the case of
antiferromagnetic exchange interactions.

As a general principle (but with notable exceptions, depending mainly
on the lattice parameter), the magnetic ground state of TM pnictides
and chalcogenides in the zinc-blende structure is ferromagnetic, if
the total number of valence electrons $Z_{\mathrm{tot}}\leq
12$. According to eq.~\ref{eq:2}, this means that the maximum moment
in the half-metallic state is 4~\muB. Thus, for instance, MnAs is
expected to be ferromagnetic, but MnSe or FeAs are expected to be
antiferromagnetic (or non-collinear). The reasoning behind this
rule is the following. Ferromagnetic order in these compounds is
brought about by the double exchange mechanism (energy is gained by a
band-broadening at \EF\ due to hybridizations) involving the spin-up
antibonding \tg\ states. Once $Z_{\mathrm{tot}}$ exceeds 12, the
spin-up \tg\ states are fully occupied, and the double exchange
mechanism is no more present (the band broadening does not bring any
gain in energy); thus the ferromagnetic order is lost.\footnote{The
same cause of ferromagnetism has been identified in some TM-doped
diluted magnetic semiconductors~\cite{Akai98}, although those systems
are rather more complicated involving also a Zener $p$-$d$ exchange.}

\subsection{Total energy results}

In accordance with the above, total energy
calculations~\cite{Shirai01} have shown that VAs, CrAs and MnAs
are ferromagnetic, while FeAs was found to have an
antiferromagnetic ground state (the possibility of a non-collinear
state was not examined in~\cite{Shirai01}). The energy gain for
the formation of the ferromagnetic state (compared to the
antiferromagnetic) is about 0.2~eV for VAs, 0.3~eV for CrAs, and
0.1~eV for MnAs; the antiferromagnetic state of FeAs provides an
energy gain of 0.1~eV compared to the ferromagnetic state (all
energies calculated within the LSDA)~\cite{Shirai01}. Calculations
on MnTe in the zinc-blende structure have shown that it also has
an antiferromagnetic ground state~\cite{Wei87}.  Other
calculations have confirmed this aspect. Total energy calculations
have confirmed a ferromagnetic ground state for
MnAs~\cite{Zhao02}, CrAs~\cite{Shirai03,Sanyal03} and
CrSb~\cite{Shirai03,Sanyal03}.

A recent study~\cite{Sanyal06} on MnAs concludes that, as \EF\ enters
the spin-down conduction band, there occurs a Fermi surface nesting
(i.e., a large area of the spin-down Fermi surface is almost parallel
to the spin-up Fermi surface, being separated by an almost constant
vector $\vec{q}$), allowing for virtual spin-flip excitations of
wavevector $\vec{q}$ and leading to a corresponding non-collinear
state. From another viewpoint, it has been
suggested~\cite{Kim04,Asada93} that there are two factors determining
the first-neighbour Mn-Mn interaction in MnAs: a direct
Mn($d$)-Mn($d$), antiferromagnetic interaction and an indirect
ferromagnetic Mn($d$)-As($p$)-Mn($d$) interaction. The former is
dominant at close Mn-Mn distances, inducing an antiferromagnetic
state, but as the lattice parameter increases the overlap of the $d$
states of neighbouring Mn atoms is lost, so that the indirect,
ferromagnetic interaction dominates. This fragility of ferromagnetism
in MnAs is in accordance to the results of exchange interaction
studies (section \ref{sec:exc}).

\subsection{Exchange interactions in the ground state and calculations
  of $T_C$\label{sec:exc}}

Sakuma~\cite{Sakuma02} reports a systematic investigation of the
interatomic exchange interactions in CrAs, MnSi, MnGe, and MnSn in the
zinc-blende structure.  Under the assumption of the validity of a
Heisenberg model, with a Hamiltonian of the form
\begin{equation}
H=\sum_{ij}J_{ij}\,\hat{e}_i \hat{e}_j
\end{equation}
(where $\hat{e}_i$ is the magnetic moment direction at site $i$), the
exchange constants $J_{ij}$ are found via the Liechtenstein
formula~\cite{Liechtenstein87}.  In this formula, the {\it ab-initio}
band structure properties enter, so that the exchange constants are
calculated without any adjustable parameter.  The coefficient
$J_0=\sum_{i\neq 0}J_{0i}$, corresponding to the band-energy cost for
flipping the magnetic moment of a single atom, reflects a
``single-site spin stiffness'', and the ferromagnetic state is stable
when $J_0>0$; a rotation of the local moment by a small angle
$\delta\theta$ costs energy $\delta E = J_0 (1-\cos\delta\theta)$,
while the mean-field Curie temperature is
$T_C^{\mathrm{mf}}=2J_0/3k_B$ (with $k_B$ the Boltzmann constant).
This method is not as accurate as a full self-consistent calculation
of a magnetic moment flip, but it is much faster and efficient in
showing trends.

All studied compounds in~\cite{Sakuma02} were found to be
ferromagnetic for a wide range of lattice parameters; $J_0$ increases
as a function of lattice parameter, as \EF\ enters deeper into the
half-metallic gap. Furthermore, by treating \EF\ as a parameter for a
fixed band structure (and calculating $J_0(E_F)$), the relative
contribution to the coupling at each energy was found as a function of
band-filling. It was shown that the spin-up antibonding \tg\ states
contribute to the ferromagnetic stability, while the spin-down
conduction band states (\eg\ and antibonding \tg) induce an
antiferromagnetic coupling and cause a change of sign to $J_0$ as \EF\
enters the conduction band. The highest values for $J_0$ are found for
\EF\ in the middle of the gap. The local-moment-flip energy cost in
CrAs, in the GaAs lattice constant (and at the real \EF), was found to
be about 0.3~eV, which compares well with the
ferromagnetic-antiferromagnetic energy difference of 0.3~eV calculated
by Shirai~\cite{Shirai01}.

Interatomic exchange interactions were also studied in \cite{Sanyal03}
for V, Cr, and Mn compounds with As, Sb, and P. It was found that the
strongest tendency for ferromagnetism is among the Cr-based compounds,
a result verified in~\cite{Sasioglu05} for pnictides. In CrAs, the
first-neighbour Cr-Cr exchange constant $J_1$ is dominant, of the
order of 15~meV. The second-neighbor Cr-Cr interaction $J_2$ is
negative (antiferromagnetic) and an order of magnitude smaller (about
$-0.1$~meV); due to the half-metallic gap, the exchange constants fall
off rapidly with distance.

In reference~\cite{Sasioglu05}, the stability of ferromagnetism in
zinc-blende CrAs, CrSe, MnAs, MnC, MnSi, and MnGe was studied,
again based on the calculation of exchange constants. Similarly to
reference~\cite{Sakuma02}, lattice compression was found to
destabilize the ferromagnetic state, as \EF\ approaches and enters
the spin-down conduction band. The effect was found to be quite
drastic in MnAs, with the ferromagnetic Mn-Mn interaction $J_1$
decreasing in magnitude, and the antiferromagnetic Mn-Mn
interaction $J_2$ increasing in magnitude; this effect was even
stronger in CrSe. (In the other compounds, the trends are similar,
but not as drastic.) Thus, as the MnAs lattice parameter is
reduced from 5.87~\AA\ to 5.68~\AA\ (a change of only about
3.5\%), the Curie temperature (calculated within the random phase
approximation) $T_C^{\mathrm{rpa}}$ drops from 551~K to 136~K; and
CrSe, for a similar compression, is found to change phase from
ferromagnetic to antiferromagnetic ($T_C^{\mathrm{rpa}}$ becomes
negative). The trend of the mean-field result,
$T_C^{\mathrm{mf}}$, is similar. An inspection of the density of
states leads to the conclusion that the antiferromagnetic
susceptibility becomes dominant as \EF\ enters the spin-down
conduction band due to compression; this conclusion is in
agreement with the result of Sakuma~\cite{Sakuma02}, as well as
with the results reported by K\"ubler~\cite{Kubler03} and by
Sanyal et al.~\cite{Sanyal06}.

Curie temperatures were calculated in references~\cite{Sanyal03},
\cite{Sasioglu05} and \cite{Kubler03}. To this purpose, several
methods have been employed. For the solution of the Heisenberg
model (after calculation of the exchange constants by
Brillouin-zone integration of static-spin-spiral energies), there
have been used the Monte-Carlo method~\cite{Sanyal03} (which are
the most accurate), the mean-field approximation (mfa)
\cite{Sanyal03,Sasioglu05} (overestimating $T_C$), or the
random-phase approximation~\cite{Sasioglu05} (rpa) (more accurate
than the mfa). Alternatively, the approach in \cite{Kubler03}
identifies the non-local dynamic susceptibility including an
adjusted damping parameter. Due to the different approaches in the
various works, and without experimental results on $T_C$ in bulk
systems, a direct comparison of the result for $T_C$ is
meaningless, however a comparison of $T_C$-trends calculated
within the same method is definitely meaningful. It is a common
conclusion that the ferromagnetic state is most robust in CrAs,
which shows the highest calculated value of $T_C$ (ranging between
790~K and about 1200~K, depending on the method).

\section{Stability of the zinc-blende phase\label{sec:6}}

The bottleneck for the fabrication of chalcogenides and pnictides in
the zinc-blende structure seems to be their structural
instability. The ground state structure of these compounds is in many
cases the hexagonal NiAs structure (MnP structure for CrAs), in which
they are not half-metallic. The zinc-blende phase is metastable, and
needs therefore to be stabilized via pseudomorphic growth on a
semiconductor substrate of the zinc-blende or diamond structure.

Total energy calculations
\cite{Liu03,Pask03,Xie03,Xie03b,Zhang03,Zheng04,Zhao02,Sanvito00,Continenza01,Miao05,Zhao05}
have confirmed that the NiAs structure has considerably lower
energy than the zb structure for almost all TM pnictides and
chalcogenides, with the exception of nitrides
\cite{Miao05b,Miao05e} which are reported to have a rock-salt
ground state (for the early transition metal compounds ScN-CrN) or
a zinc-blende ground state (for the late transition metal
compounds, MnN-CoN); the nitrides, however, have too small a
lattice parameter to be half metallic, and are beyond the scope of
our discussion. The bulk zinc-blende phase has typically a higher
energy of the order of $\Delta_{\mathrm{bulk}}\sim 0.3$--1~eV per
formula unit than the NiAs phase (when both are calculated at
their corresponding equilibrium lattice constants); the NiAs phase
has also a considerably smaller equilibrium unit cell volume
$V_{\mathrm{eq}}$ (e.g., reference~\cite{Sanvito00} reports for
MnAs in the NiAs structure $V_{\mathrm{eq}}\approx 31$~\AA$^3$ and
$V_{\mathrm{eq}}\approx 45$~\AA$^3$ in the zinc-blende structure).
It has been reported \cite{Xie03} that $\Delta_{\mathrm{bulk}}$ is
lower for chalcogenides (CrTe, CrSe, and VTe were considered, with
$\Delta_{\mathrm{bulk}}\sim$0.3 to 0.5 eV) than for pnictides
($\Delta_{\mathrm{bulk}}\sim$0.5 to 1.0 eV), possibly giving an
advantage to the growth of chalcogenides in the zinc-blende phase.

Typically, under sufficient volume expansion, the NiAs phase
becomes unfavourable compared to the zinc-blende phase (the energy
curves cross, and $\Delta_{\mathrm{bulk}}$ becomes negative at
some increased lattice constant). However, more important than
$\Delta_{\mathrm{bulk}}$ is the epitaxial energy difference
$\Delta_{\mathrm{epi}}$ (i.e., the energy difference between the
two phases at a particular lattice constant when also a change of
the c/a ratio is allowed), since a compound growing epitaxially on
a substrate of increased lattice parameter will be able to adjust
its c/a ratio in favour of elastic energy. This was pointed out in
reference~\cite{Zhao05}, where a number of compounds were studied
in this aspect (MnAs, CrAs, CrSb, CrS, CrSe, and CrTe). It was
found that $\Delta_{\mathrm{epi}}$ always favours the NiAs phase
(for all hypothetical substrate lattice parameters), except for
CrSe, for which $\Delta_{\mathrm{epi}}$ becomes negative at
approximately 6.2~\AA, favouring an epitaxial growth in the
zinc-blende phase for substrates of higher lattice constant.
Nevertheless, these conclusions are valid for the growth of thick
layers, while a few monolayers will be much affected by the
interfacial energy. This agrees with the experimental observation
that CrAs and CrSb can be grown in the zinc-blende phase for
thicknesses up to a few monolayers, and then break up.

The stability of the zinc-blende phase has also been examined with
respect to tetragonal and rhombohedral
deformation~\cite{Xie03,Zheng04,Yamana04,Shi05,Miao05d} (starting from
the equilibrium volume in the zinc-blende structure). It was found
that, under rhombohedral deformation, chalcogenides are stable
~\cite{Xie03,Shi05}, while, among the studied pnictides, only CrAs,
CrSb, and MnAs are stable (although rather soft, with shear moduli of
the order of 3-5~GPa)~\cite{Shi05}. Tetragonal deformation was found
to be favourable for MnSb and MnBi \cite{Zheng04}, with an equilibrium
c/a$\sim 0.75$, but the half-metallic character was not changed. This
stability of the half-metallic character under tetragonal distortion
in many TM pnictides and chalcogenides is the conclusion of a number
of works~\cite{Xie03,Zheng04,Yamana04,Shi05,Mavropoulos04b,Miao05c}.
In particular, a tetragonalization along the [001] direction (expected
in the presence of a (001) interface) distinguishes the $z$ axis and
causes a splitting of the \tg\ and \eg\ irreducible representations at
the Brillouin zone center $\Gamma$~\cite{Mavropoulos04b}. The \tg\
threefold degeneracy splits into a twofold degeneracy (involving the
$d_{xz}$-$p_y$ and $d_{yz}$-$p_x$ hybrids) and a non-degenerate level
(involving the $d_{xy}$-$p_z$ hybrids); the \eg\ twofold degeneracy
splits into two non-degenerate levels, one involving the $d_{z^2}$
levels and one involving the $d_{x^2-y^2}$ levels. These splittings
affect the spectral function at $\Gamma$, but they are in no way
important for the overall DOS and the gap.

However, in reference~\cite{Miao05d} it is reported that a number
of compounds (including CrSe and MnAs) cannot be half-metallic
even at large the substrate lattice constants, because of the
unfavourable vertical relaxation and c/a ratio. An interesting
conclusion of the same work is that, for some pnictides (including
CrAs, MnAs, and MnSb) the zinc-blende phase is unstable with
respect to a ``gliding'' tetragonal deformation, where the
in-plane lattice parameter changes with simultaneous change of the
c/a (so that the bond length is kept constant). In practice,
however, a tetragonal deformation of this particular kind cannot
take place, because the in-plane lattice parameter is fixed by the
substrate.

\section{Surface half-metallicity\label{sec:4}}

The surfaces of half-metallic ferromagnets are not guaranteed to
be also half-metallic. This is because wavefunction hybridizations
and bonding-antibonding splittings, essential to the appearance of
the gap, change significantly at the surface; the ``missing
neighbours'' can result in the appearance of surface states in the
gap, destroying half-metallicity. This is, for instance, the case
in most surfaces of half-metallic Heusler alloys
\cite{Heuslers-Surf}. Although half-metallic surfaces are not of
particular technological interest (contrary to half-metallic
interfaces), they are of value for proving the half-metallic
property by spin-polarized photoemission experiments, which are
surface sensitive\footnote{See, e.g.,
reference~\protect{\cite{Rader05}} on an experimental study on the
(0001) surface of MnSb in the (non-half-metallic) NiAs structure}.
In zinc-blende pnictides and chalcogenides, it was demonstrated by
{\it ab-initio}
calculations~\cite{Galanakis03,Galanakis02,Kang05,Byun06} that the
(001) surfaces can, conditionally, be half-metallic.

Surface half-metallicity in the (001) surfaces of these compounds
can occur only if they are terminated with the TM.  In case of
termination with the $sp$ atom, dangling bonds appear within the
half-metallic gap, destroying half-metallicity (the possibility of
surface reconstruction was, however, not considered in the
calculations in~\cite{Galanakis03,Galanakis02}). Similar effects
were found in reference~\cite{Qian05}.

In case of TM termination, the electronic structure and DOS changes at
the surface (even if it is half-metallic). A simple electron counting
shows that, due to the missing $sp$ neighbour, $(8-Z_{sp})/2$
additional electrons must be accommodated by the surface TM, where
$Z_{sp}$ is the valency of the $sp$ atom. To understand this, consider
that, in the bulk, these electrons would occupy bonding $p$-$d$
hybrids; here, half of these hybrids are missing. Local charge
neutrality requires that these electrons remain in the vicinity of the
surface (mostly in the surface layer). Moreover, half-metallicity is
only preserved if they occupy spin-up states (otherwise \EF\ will
enter the conduction band). Therefore, in the cases when
half-metallicity is preserved, the spin moment at the surface
increases by $(8-Z_{sp})/2$~\muB\ per surface atom. This also leads to
an increase of the exchange splitting, shifting the spin-down \eg\
levels higher in energy, thus also increasing the gap width. Such is
the calculated case in, e.g., CrAs, VAs, or CrSe. On the other hand,
MnAs has a bulk moment of 4~\muB\ per formula unit. If its
Mn-terminated, (001) surface were to retain the half-metallic
character, the local moment should increase to 5.5~\muB. This,
however, is not possible, since the spin-up \eg\ and antibonding \tg\
states can accommodate at most 5 electrons, leading to a maximum moment
of 5~\muB. As a result, \eg\ spin-down states are occupied, and
half-metallicity is lost.

Also the (110) surfaces of CrP and CrAs were found to be
half-metallic in reference~\cite{Lee06,Lee06b} (the calculation of
CrP~\cite{Lee06} was done at an expanded lattice constant
(5.48\AA), at which also bulk half-metallicity appears).
Interestingly, the (110) surface contains both kinds of atoms (Cr
and P or Cr and As), and no dangling bonds appear to destroy the
half-metallic property.

Since, from the electronic structure point of view, all ferromagnetic
zinc-blende pnictides and chalcogenides are similar, it is expected
that surface half-metallicity is likely to appear in such
compounds. However, more theoretical investigations are necessary to
verify this, examining in particular the effect of lattice relaxations
and surface reconstruction.

\section{Interfaces with semiconductors\label{sec:5}}

For technological applications of half-metals, especially in
magnetic tunnel junctions, it is imperative to eliminate the
interface states in the gap region (i.e., for spin down) of the
half-metal-semiconductor interface. The reason is that these
states can act as carrier reservoirs and contribute to the
transport, although they are localized, because the tunneling rate
can be much lower than the refill rate of these states by
spin-orbit coupling or inelastic effects~\cite{Mavropoulos05}.
Contacts of half-metallic Heusler alloys with semiconductors are
known to host such interface states in almost all studied cases
\cite{Heuslers-Interf}. It is therefore gratifying that pnictides
and chalcogenides in the zinc-blende structure retain their
half-metallicity also at the interfaces with semiconductors (which
are anyhow the natural substrate for growth of these half-metals).
The reason is that the bonding character changes coherently at the
interface. For example, at a CrAs/GaAs interface (with an As
interface layer), the $p$ states of the As interface atoms
hybridize with the Cr $d$-states on the one side and with the Ga
$s$ states on the other. Thus, the bonding-antibonding splitting
does not cease at the interface. Such a behaviour is expected also
in view of half-metallicity in TM-doped diluted magnetic
semiconductors~\cite{Akai98}.

Interfaces of half-metallic pnictides and chalcogenides in the
zinc-blende structure with III-V and II-VI semiconductors (in the same
structure) were studied theoretically in
\cite{Mavropoulos04b,Qian05,Shirai04,Nagao04,Bengone04,Cha04,Fong04b,Wu06}.
It was found that no interface states appear at \EF\ within the
half-metallic gap, because of the coherent bonding described
above. Moreover, moderate tetragonalization (which is expected in the
case of a slight lattice mismatch) was not found to alter the
half-metallic character. (See section \ref{sec:6} for a discussion of
tetragonalization).

The half-metallic gap was found to be persistent also in the case of a
50\% itermixing at the interfaces \cite{Mavropoulos04b}, in the sence
that, at CrAs/GaAs for instance, the interface layer is half occupied
by Ga and half by Cr. This is expected, since, again, the bonding
continues coherently, and it is known that half-metallicity is present
for a wide range of concentrations in diluted magnetic
semiconductors~\cite{Akai98}.

The gap remains present at the interface also in the case that the
semiconductor and half-metal anions are different (e.g.,
CrAs/GaSb). If, however, the valency of the cation changes, the
electronic structure is more complicated \cite{Mavropoulos04b}. As an
example we take a CrSb/ZnTe(001) interface, where ZnTe is a II-VI
semiconductor, while Sb is a group V element. Interface states appear
at \EF\ in the case of direct Zn-Sb contact, i.e., if the layered
structure is of the form ...CrSbCrSbZnTeZnTe... On the other hand, if
the interface atoms in contact are Cr and Te (structure of the form
...CrSbCrTeZnTe...), the gap is preserved at the interface. In the
latter case, there is also a local increase of magnetic moment at the
interface Cr atoms from 3 to 3.5 \muB\ because the Te atoms have one
less $p$ hole than the Sb atoms, thus the interface Cr spin-up charge
is increased.

\subsection{Delta-doped zinc-blende semiconductors\label{sec:delta}}

Delta-doping of semiconductors by TM constitutes a special,
limiting case of half-metal-semiconductor interface
\cite{Shirai98,Kawakami00,Qian06,Sanvito01}. Such ``digital''
compounds are of special interest, because of their
two-dimensional half-metallic behaviour and the highly anisotropic
transport properties~\cite{Sanvito01}. Also, delta-doping is a
possible way to overcome the low solubility limit of transition
metals in III-V semiconductors, making such constructions
interesting as a special case of diluted magnetic semiconductors.

Experiments on GaAs, delta-doped with Mn (in the submonolayer
range), found the compound to be ferromagnetic \cite{Kawakami00}.
The Curie temperature was dropping with the distance between the
Mn delta-layers, but saturated for large distances, indicating
ferromagnetism of each single layer. A two-dimensional
ferromagnetic, half-metallic character was also found by
calculations in GaAs delta-doped with Mn
\cite{Shirai98,Sanvito01}. In reference~\cite{Sanvito01}, the
conductivity was also calculated, and it was found that the
in-plane conductivity was metallic (only for spin up), with the
current confined at and close to the Mn layers, while the
perpendicular-to-the-plane conductivity was extremely low,
reflecting the tunneling between the Mn layers.  It is known that
ferromagnetic order in DMS (in particular Ga$_{1-x}$Mn$_x$As) is
assisted by the $p$ holes introduced by the Mn dopants (due to the
$p$-$d$ repulsion)~\cite{Akai98}. As Mn-delta-doped GaAs can be
viewed as a limiting case of DMS, it is interesting to see
\cite{Sanvito01} that also the $p$-holes are in this case confined
at and close to the Mn layers.  Calculations on a delta-doped
layer of Mn in Si \cite{Qian06} and Ge \cite{Continenza04} also
showed two-dimensional half-metallic ferromagnetism, based on the
same basic mechanism.

In reference~\cite{Kim04}, a MnAs/Si superlattice is calculated to
induce antiferromagnetic order to MnAs, possibly due to the small
lattice parameter; a small 2\% tetragonal distortion (with
increased c/a) restores the ferromagnetic state.

Concerning the exchange interactions, a study of the exchange
constants in Mn-delta-doped Ge and GaAs was presented in
reference~\cite{Picozzi06}. It was found that the systems are
ferromagnetic, dominated by $J_1>0$ \footnote{For the notation see
section~\ref{sec:exc}}, while $J_2<0$ is lower in magnitude and
antiferromagnetic (as is the case in the bulk of zb compounds MnGe
and MnAs~\cite{Sasioglu05}). Since density-functional calculations
underestimate the semiconductor gap, calculations were performed
with an ad-hoc increased gap, in order to examine its effect on
the exchange constants. On increasing the semiconductor gap
(either for Ge or for GaAs), the calculations~\cite{Picozzi06}
showed the systems to become more ``ferromagnetic'', in the sense
that $J_1$ increases while $J_2$ decreases in magnitude.

\subsection{{\it Ab-initio} interface engineering}

The ferromagnetic and antiferromagnetic properties of TM
chalcogenides and pnictides can be applied to novel tunnel
junctions. In reference~\cite{Mavropoulos05}, first principles
calculations showed how one can design a tunnel junction of
half-metallic CrTe elements which are antiferromagnetically
coupled via a semiconducting CdTe spacer. The idea is to introduce
a monolayer Mn at the CrTe/CdTe interface, so that the junction is
of the form ...(CrTe)$_2$MnTe-(CdTe)$_n$-MnTe(CrTe)$_2$... along
the [001] direction. Since the Mn-Mn interaction in CdTe is known
to be antiferromagnetic, the two magnetic parts of the junction
are antiferromagnetically coupled via this interface engineering.
The Mn layer couples antiferromagnetically also to the CrTe
layers, still resulting in half-metallic parts. The CdTe thickness
can be varied so that the interaction energy is tuned to a
desirable low level. In the ground state, the junction is
insulating, since the spin-up states at \EF\ of the two
half-metallic parts are in opposite spin directions. Application
of an external magnetic field can orient the moments of both
half-metallic parts in parallel, so that electrons can tunnel
between the spin-up bands of the half-metallic parts, switching on
the conductance. Since such a junction is free of interface states
at \EF\ in the half-metallic gap, its realization would constitute
an ideal half-metal-semiconductor switch.

A recent investigation~\cite{Nakamura06} examined CrSe/MnSe and
CrTe/MnTe ferromagnetic/antiferromagnetic (001) interfaces. Since
MnSe and MnTe are antiferromagnetic semiconductors, while CrSe and
CrTe are ferromagnets, an exchange bias is formed at the
interface. The antiferromagnetism was reported to be
layer-by-layer in the \{001\} direction. It was found that, in the
case of an uncompensated antiferromagnetic (001) interface (where
the antiferromagnetic layers are perpendicular to the interface
plane), a collinear half-metallic magnetic state is formed
throughout the junctions, with an exchange bias of $\Delta
E=-0.41$~eV/a$^2$ for CrSe/MnSe and $-0.41$~eV/a$^2$ for CrTe/MnTe
(here, a is the in-plane lattice parameter). In the compensated
(001) interface (where the antiferromagnetic layers are parallel
to the interface plane), a spin-flop (non-collinear) ground state
was found, with the Cr moments being perpendicular to the Mn
moments; the corresponding exchange bias is $\Delta
E=-0.08$~eV$/a^2$ for CrSe/MnSe and $-0.03$~eV$/a^2$ for
CrTe/MnTe. A similar spin-flop sate was also calculated in
reference~\cite{Nakamura05} in (CrSe)$_1$(MnSe)$_1$ and
(CrTe)/(MnTe) multilayers. In view of the fragile
ferromagnetism~\cite{Sasioglu05} and
half-metallicity~\cite{Miao05d} of CrSe, we consider these
conclusions rather more important for CrTe/MnTe than for
CrSe/MnSe.

\section{Effect of disorder\label{sec:7}}

The half-metallic property of zb TM-pnictides and chalcogenides is
based on the hybridization gap formed by the $d$ states of the TM and
the $p$ states of the $sp$ atom, and on the large exchange splitting
pushing the spin-down \eg\ states high in energy. Therefore, disorder
(e.g., in the form of antisites or swaps) is expected to result to a
loss of half-metallicity, since the $p$-$d$ hybridization is expected
to change at an impurity, and the magnetic moment of an impurity can
be lower than the moment of the TM.

The energetics of Cr-As swaps in zb-CrAs have been calculated from
first principles in reference~\cite{Shirai04b} for swaps up to
10\%. Three magnetic states were considered for the
Cr$_{\mathrm{As}}$ antisites (Cr at the As position):
ferromagnetic, antiferromagnetic, and spin-glass. It was found
that the Cr-As swap is not favoured energetically in any of these
cases. However, if such a swap occurs, it was found that the Cr
antisites are antiferromagnetically coupled to the rest of the Cr
atoms (leading to a ferrimagnetic overall picture), with an energy
gain of about 2~eV per Cr antisite compared to the ferromagnetic
swap-state; the spin-glass energy was still higher. Furthermore,
it was found that the Cr antisites introduce states at \EF\ within
the half-metallic gap, destroying the half-metallic property.

In another study~\cite{Bengone04}, the mixing energy of As antisites
in zb (Cr$_{1-x}$As$_x$)As was calculated, with respect to the phase
separated zb CrAs and pure As. It was found that, if As is in the
ground-state trigonal structure ($\alpha$-As), phase separation is
favoured leading to clean zb CrAs. If, however, As is in the diamond
structure (which could be stabilized by a GaAs substrate), the
formation of As$_{\mathrm{Cr}}$ antisites is favoured for $x\approx
0.6$. In the presence of As$_{\mathrm{Cr}}$ antisites, it was found
that spin-down states are introduced at \EF, and half-metallicity is
lost.

Furthermore, the effect of antisites on the density of states and
on the magnetic moments was studied in
reference~\cite{Galanakis06} for CrAs, CrSb, CrSe, CrTe, VAs, and
MnAs. Alloys of the type Cr$_{1+x}$As$_{1-x}$, for $-0.5<x<0.5$,
(and similarly for the other compounds) were calculated. Notably,
it was found that half-metallicity is possible in some disordered
cases. Mn$_{1+x}$As$_{1-x}$, for $x>0$, remained half-metallic,
with an increase of the magnetic moment (the Mn$_{\mathrm{As}}$
antisites were found to align ferromagnetically to the rest of the
Mn atoms). Also in all Cr compounds, for $x>0$, the half-metallic
character is kept (for not-too-high values of $x$), while the
magnetic moment dropped (the Cr antisites were found to align
antiferromagnetically to the rest of the Cr atoms).

\section{Other tetrahedrally bonded structures\label{sec:8}}

\subsection{TM-based wurtzite compounds}

The tetrahedral environment in zb TM pnictides and chalcogenides
is the essential factor to the formation of $p$-$d$ hybrids and
the gap. Therefore, one expects that TM pnictides and
chalcogenides in the wurtzite structure (wz), which provides an
almost tetrahedral environment, should also be half-metallic.
Based on these ideas, Xie {\it et al.}~\cite{Xie03c} made a
systematic theoretical study of such wz compounds, including MnSb,
CrAs, CrSb, VAs, VSb, CrSe, CrTe, VSe, VTe. They concluded that
half-metallicity should be present also for the wz structure. The
density of states is very similar to the one in the zb structure;
in the energy bands there is a degeneracy lifting at the Brillouin
zone center, since there is no tetrahedral symmetry (the
environment of each atom is almost, but not exactly, tetrahedral).
The total moments follow the ``rule of 8'' (eq.~(\ref{eq:2})).
Half-metallicity in wz CrS, CrSe, CrTe was also reported in
reference~\cite{Zhang04c}. Bismuth-based compounds in the wz
structure were studied in reference~\cite{Zhang05}. VBi and CrBi
were found to be half-metallic at their respective equilibrium
lattice parameters (neglecting the strong spin-orbit coupling
caused by Bi), while MnBi was found not to be half-metallic; a
lattice expansion was needed to restore half-metallicity.

Total energy calculations~\cite{Miao05} reveal that the energy of the
TM pnictides wz structure is very close to the energy of the zb
structure, and the lattice constants of the two phases are almost
equal. This means that the wz structure is also metastable.

\subsection{Nanoclusters and nanodots}

Nanoclusters of zb TM-pnictides and chalcogenides have also been
studied theoretically~\cite{Nakao04,Qian04}. This is interesting
in view of the experimental results on MnAs
nanodots~\cite{Oshima02,Ono02,Okabayashi04}, although the
calculated nanoclusters are much smaller, not exceeding a size of
50 atoms (the experimental nanodots were 16~nm in diameter).
Reference~\cite{Nakao04} presents results on free-standing
nanoclusters TM$_{13}$X$_{16}$ with TM=(V, Cr,Mn) and
X=(N,P,As,Sb,S,Se,Te). By changing the lattice constant, several
ferromagnetic and antiferromagnetic magnetic configurations were
arrived at (although no total energy results were reported). The
$sp$ atoms at the cluster boundary were found to produce dangling
bonds, affecting in some cases the half-metallic character. When
CrAs nanoclusters were embedded in a GaAs matrix, the dangling
bonds were found to vanish, due to the bonding of the boundary
atoms with the matrix. In reference~\cite{Qian04} the focus was in
a free-standing zb MnAs nanocluster containing 41 atoms of which
13 were Mn atoms (including a central Mn atom and a first shell of
12 Mn atoms). Structural relaxation was taken into account. It was
found that, in the ground state, the central atom couples
antiferromagnetically to the first shell. The ferromagnetic state
was 125~meV higher, and half-metallic, with a gap of 1.83~eV.

\section{Recent results on the role of electron correlations and on
  the temperature dependence of the polarization\label{sec:12}}

Density-functional calculations, within the LSDA or GGA, can capture
only a part of the electron exchange and correlation effects. In
particular dynamic effects need to be described by more elaborate (and
computationally heavier) extensions, such as the LSDA+DMFT (dynamical
mean-field theory).

An interesting effect of dynamic electron correlations on
half-metallic ferromagnets is the appearance of non-quasiparticle
states (NQS) in the half-metallic gap just above \EF. These
states, originating from virtual electron-magnon scattering
processes, are captured within the DMFT and were calculated in
reference~\cite{Chioncel05} for the case of CrAs. The NQS appear
at characteristic magnon energies, i.e., a few tens of meV above
\EF, and can therefore affect the transport properties even at
small voltages. However, to our knowledge, their exact effect on
spin-polarized transport has not yet been clarified; such states
should be present in all half-metallic systems, because the
mechanism that brings them about is rather
general~\cite{Irkhin02}.

Another effect of dynamic correlations can be a shifting of
spectral weight towards the Fermi level, which is expected due to
the form of the real part of the self-energy $\Sigma(E)$ near \EF:
$\mathrm{Re}[\partial\Sigma/\partial E]_{E_F}<0$
\footnote{$\Sigma(E)$ represents a measure of renormalization of
the spectral density due to correlations; the Green function, of
which the poles determine the spectrum, has the form
$G(E)=(E-H^0-\Sigma(E))^{-1}$, where $H^0$ is the Hamiltonian in
the absence of electron correlations.}. This has an interesting
effect on VAs, as was calculated in reference~\cite{Chioncel06}.
Within the LSDA (or GGA), VAs is found to be a ferromagnetic
semiconductor, because \EF\ lies exactly between the spin-up \eg\
and antibonding \tg\ states (for spin-down it is in the
half-metallic gap). The spin-up gap is tiny, of the order of
50~meV. Within the LSDA+DMFT, it is found that the spin-up \eg\
and \tg\ states are shifted towards the Fermi level, closing the
gap completely and causing a transition from semiconducting to
half-metallic behaviour.

We would also like to point out the problem of absence of a Fermi
edge observed in the photoemission spectroscopy of ferromagnetic
zb MnAs nanodots~\cite{Oshima02,Ono02} (see also
section~\ref{sec:2}). This is in contrast to theoretical results
within density-functional theory, and to our knowledge its
explanation is an open problem which deserves more attention in
the future.

An open question is the behaviour of the spin polarization $P$ at \EF\
at elevated temperatures $T$, since it is obvious that $P$ must drop
at high $T$. Only few theoretical attempts have been made in this
direction~\cite{Itoh00,Dowben02,Chioncel06b,Lezaic06}, examining
different mechanisms that can contribute to the drop of $P(T)$. These
depend on many factors, including the exact band structure and
position of \EF\ in the gap, but the theoretical methods have not been
applied to zb pnictides and chalcogenides. In order to obtain a
quantitative result, these theories have to be refined and unified (in
our opinion, this is one of the most challenging theoretical problems
in the field of half-metallic ferromagnets); they indicate, however,
that a high $P$ at room temperature requires (i) $T_C$ to be
considerably higher than room temperature and (ii) \EF\ to be deep in
the gap, and not in the proximity of a band edge. Experimentally,
$P(T)$ can be approached indirectly via transport measurements in
magnetic tunnel junctions under some assumptions on the
polarization-dependence of the magnetoresistance~\cite{Bowen04}.

\section{Summary and outlook\label{sec:13}}

In this review we have summarized the work reported on transition
metal pnictides and chalcogenides in the zinc-blende structure.
Relatively few experimental results have been published. We
believe that this is due to the difficulties in experimental
growth of these structures. So far, experimental fabrication has
been demonstrated only for ultrathin layers (including
multilayers) of CrAs and CrSb and nanodots of MnAs, and evidence
exists for ultrathin layers of MnAs and CrTe (see section
\ref{sec:2}). Magnetic measurements have been reported, but no
evidence has been shown of half-metallic behaviour (or of its
absence, except for the insulating phase reported for zb MnAs
nanodots \cite{Oshima02,Ono02}). At the same time, there are many
theoretical and computational investigations on the subject,
including other related structures. To a great extent, these
concern the structural stability of the zb phase (see
section~\ref{sec:6}), and demonstrate that this phase can at most
be grown epitaxially for a few layers (in agreement with
experiments on CrAs and CrSb). Although no prediction on future
experimental results can be made, we conclude that, if zb
pnictides or chalcogenides are to be used in spintronics
applications (such as magnetic tunnel junctions or spin
injection), it has to be done in a way that a few monolayers are
sufficient.

On the other hand, from the calculated electronic structure and
transport point of view, these compounds are excellent, and growth of a
few monolayers is probably enough for applications. Theoretical
investigations show that, at the contact with semiconductors, no
interface states in the gap are present, and that the half-metallic
state is also robust against tetragonal distortions.  The
ferromagnetic interactions are dominant even in delta-doped (i.e.,
one-monolayer) cases, so that even one monolayer should be enough to
provide fully spin-polarized transport.  Calculations including
dynamic correlations show that the half-metallic property (at least in
CrAs) is not affected, except for the non-quasiparticle states just
above \EF.

The material with the most robust ferromagnetism is probably CrAs,
which is found by calculations to have the highest Curie temperature.
It is reasonable to assume that, among the materials which were
reviewed here, the best candidates for future applications are CrAs
and CrSb (since they were the zb pnictides fabricated so-far), or
delta-doped compounds (including Si and Ge) in the monolayer or
sub-monolayer range.

\end{document}